\documentclass[aps,prd,nofootinbib]{revtex4}
\usepackage[colorlinks=true, pdfstartview=FitV, linkcolor=magenta, citecolor=blue, urlcolor=magenta]{hyperref}
\usepackage[dvipsnames]{xcolor}
\usepackage{graphicx}
\usepackage{latexsym}
\usepackage{amsmath}
\usepackage{amsfonts}
\usepackage{amssymb}
\usepackage{verbatim}
\usepackage{braket}
\usepackage{extarrows}
\usepackage{tensor}
\usepackage [autostyle, english = american]{csquotes}
\MakeOuterQuote{"}

\def\f{\phi}

\newcommand{\be}{\begin{equation}}
\newcommand{\ee}{\end{equation}}
\newcommand{\bea}{\begin{eqnarray}}
\newcommand{\eea}{\end{eqnarray}}

\newcommand{\ben}{\begin{eqnarray}}
\newcommand{\een}{\end{eqnarray}}

\begin{document}

\title{Lightcone fluctuations in a five dimensional Kaluza-Klein model and an estimation on the size of the extra dimension}

\author{$^{1}$Giulia Aleixo}
\email{giulia.aleixo@academico.ufpb.br}

\author{$^{1}$Herondy Mota}
\email{hmota@fisica.ufpb.br}

\affiliation{$^{1}$Departamento de F\' isica, Universidade Federal da Para\' iba,\\  Caixa Postal 5008, Jo\~ ao Pessoa, Para\' iba, Brazil.}


\begin{abstract}
In this work we consider lightcone fluctuation effects in a five-dimensional spacetime arising as a consequence of the compactification of the extra dimension via a quasiperiodic condition, characterized by a phase angle $2\pi\alpha$. By considering light propagating in a non-compactified direction we are able to compute both the renormalized graviton two-point function and the flight time of a photon caused by the fluctuations. We show that the resulting expressions depend on the quasiperiodic parameter, the compactification length and also on the distance traveled by the photon. Based on the Near-Infrared Spectrograph (NIRSpec) sensitivity built on the James Webb Space Telescope, we discuss the possibility of making estimations on the size of the extra dimension if one assumes that the mean deviation on the flight time of photons can be observed through the NIRSpec. We also analyze the differences between the estimations in the periodic $\alpha=0$, antiperiodic $\alpha=1/2$ and other condition cases for $\alpha$.



\end{abstract}
 \maketitle


\section{Introduction}
\label{intro}
%

In a full-fledged theory of gravity divergences that commonly appear in Quantum Field Theory are supposed to be smoothed out and, as an example, we can mention divergences stemming from observables calculated exactly at the lightcone. In fact, Pauli conjectured that lightcone divergences may be removed upon quantization of gravity \cite{pauli1956}. Such possibility was subsequently discussed by Deser, Isham, and others \cite{deser1957, isham1971, isham1972}. In the context of a linearized quantum theory of gravity, for instance, it has been shown that some lightcone divergences are smeared out due to nonzero metric fluctuations \cite{ford1995}. The lightcone fluctuation effects are analogous to the Casimir effect in the sense that the presence of boundary conditions, nontrivial topology, and spacetime dimensionality can introduce perturbations to the usual quantities \cite{ford1995,ford1996,ford1999, lightconeCorda}.



These fluctuations may modify the two-point function of the quantized metric perturbation and its effects on photon propagation could, in principle, be observed: the smeared lightcone is reflected as a change in the photon speed, thereby also modifying its typical flight time \cite{ford1995, ford1996, ford1999}. This modification can vary slightly from a photon to another, however, the observed spectral lines of a light pulse will exhibit a characteristic mean deviation from the original configuration. Observable effects such as this could help us understand many phenomena involving cosmic messengers and the structure of spacetime even within a particular scope such as the linearized theory of gravity. The latter can also be applied to lightcone fluctuations in order to understand horizon fluctuations, in particular, in the case of black hole horizons, some interesting developments on black hole thermodynamics can arise \cite{ford1999}. Furthermore, since the effects of metric fluctuations resulting from modifications imposed to a particular direction are not limited to that same direction, this approach could also provide us with a powerful mean to test the existence of extra dimensions.  

After the formulation of General Relativity several attempts have been made to unify gravity and the electromagnetic theory. Among them, in 1921, Theodor Kaluza tried to grasp a solution describing a world beyond our usual $(3+1)$ dimensions introducing an extra dimension and, a few years later, Oskar Klein revisited Kaluza's work under the scope of quantum mechanics. Hence, $(n+1)$D scenarios with compact extra dimensions are often referred to as the Kaluza-Klein models, whereas the particular case of $(4+1)$ is referred to as the original Kaluza-Klein model. In recent years, such models are sought out in an attempt to describe several phenomena in many areas of physics \cite{kk_gravity, kk_largeExtraD, kk_theories, zwiebach_2004}. In particular, in Refs. \cite{HYu2000, hongweiYu2000, hongweiYu2009} lightcone fluctuations have been considered assuming periodically compactified extra dimensions. In this work we consider the $(4+1)$ dimensional Kaluza-Klein model by compactifying the extra dimension via a quasiperiodic condition, which could bring new mathematical features to our results and shed more light on the search for observable lightcone fluctuations effects and on the possible structure of the extra dimension. We also provide an estimation on the size of the extra dimension if one assumes that the flight time of a photon caused by lighcone fluctuations resides in the sensitivity range of the Near-Infrared Spectrograph (NIRSpec) on the James Webb Space Telescope \cite{ResolvingPowerNIRSPec2022}.

This work is organized as follows: In Sec. \ref{sec:LCFluc_formadeSigma} we briefly review how a linearized theory of gravity can provide a quantized perturbation $h_{\mu\nu}$ on the metric, resulting in lightcone fluctuations and, therefore, in the modification of a light pulse via the renormalized expectation value of the squared geodesic separation between two points, namely $\langle\sigma_1^2\rangle_R$. In Sec. \ref{sec:gvt_2ptFunc}, we particularize the expressions for the graviton two-point function and for $\langle\sigma_1^2\rangle_R$ to five dimensions using the plane wave expansion for $h_{\mu\nu}$. Finally, in Sec. \ref{sec:LC_KK} we obtain the exact forms of $h_{\mu\nu}$, of the graviton two-point function and of $\langle\sigma_1^2\rangle_R$ in $(4+1)$D with a quasiperiodic condition imposed on the extra dimension. In particular, in Secs. \ref{subsec:alphaNull} and \ref{subsec:alphaNonNull} we obtain the corresponding expressions for large values of $\gamma = r/\ell_c$, being $r$ the distance from the photon's source to the observer and $\ell_c$ the length of the extra dimension. We also discuss the possibility of using the NIRSpec sensitivity range to verify these results and help determine the possible structure of the extra dimension. Throughout the paper, we use natural units $\hbar = c = 1$, $G_d = (32\pi)^{-1}$.
%
\section{Lightcone fluctuations and the form of $\langle\sigma_1^2\rangle$}
\label{sec:LCFluc_formadeSigma}
Let us consider a flat spacetime described by the metric tensor $\eta_{\mu\nu}$ through which a linear perturbation $h_{\mu\nu}$ propagates. This configuration is described by the line element
\begin{equation}\begin{split}\label{lineElement_5D}
    ds^2 & = g_{\mu\nu}dx^\mu dx^\nu = (\eta_{\mu\nu} + h_{\mu\nu})dx^\mu dx^\nu \\
         & = dt^2 - d\textbf{x}^2 + h_{\mu\nu}dx^\mu dx^\nu,
\end{split}\end{equation}
where each spacetime coordinate is contained in the interval $(- \infty, \infty)$ and $\textbf{x}$ is the spatial coordinate vector. Now let $\sigma(x,x')$ be one half of the squared geodesic separation between two points $x$ and $x'$ in the spacetime described above. In the presence of a linearized metric perturbation the geodesic separation squared can be expanded in terms of the perturbation $h_{\mu\nu}$, that is
\begin{equation}\label{sigmaExpansion}
    \sigma(x,x') = \sigma_0 + \sigma_1 + \mathcal{O}(h_{\mu\nu}^2),
\end{equation}
where $\sigma_1$ is of first order in $h_{\mu\nu}$, and $2\sigma_0 (x,x') = (t-t')^2 - (\textbf{x}-\textbf{x}')^2$. We can assume the linearized perturbation to be quantized \cite{Misner1973,carroll2003spacetime}. Upon quantization, the perturbation will now act as an operator on a quantum state $\ket{\psi}$ defined such that $h_{\mu\nu}$ can be decomposed into its positive and negative frequency parts so that
\begin{equation} \label{h+h-}
    h_{\mu\nu}^+ \ket{\psi} = 0, \quad\qquad\qquad \bra{\psi} h_{\mu\nu}^- = 0.
\end{equation}
Based on the expressions above, we can say that the quantum state $\ket{\psi}$ is a vacuum state. A quick look at Eq. \eqref{h+h-} reveals that $\bra{\psi}h_{\mu\nu}\ket{\psi} = \langle h_{\mu\nu} \rangle = 0$. However, in general we have $\langle h_{\mu\nu}^2 \rangle \neq 0$, indicating that the spacetime metric $g_{\mu\nu}$ is actually fluctuating. Since $\sigma_1$ is of first order in $h_{\mu\nu}$, this same analysis applies to $\langle\sigma_1\rangle$ and to $\langle\sigma_1^2\rangle$.
Furthermore, it can be shown that the lightcone divergences are removed upon quantization of $h_{\mu\nu}$, leaving in its place a smeared oscillating lightcone, which can be understood as a fluctuation on the light speed \cite{ford1995}. This is directly connected to the nonzero values obtained for $\langle h_{\mu\nu}^2\rangle$ and $\langle\sigma_1^2\rangle$. Additionally, as it is also shown in Ref. \cite{ford1995}, the general behaviour of the smeared lightcone is not affected by taking into consideration higher orders of $h_{\mu\nu}$, therefore, we will automatically discard such contributions in our calculations. 
The effects of these fluctuations are, in principle, observable. Let us consider a source emitting evenly spaced light pulses. An observer at a distance $r$ from the source will detect a change in the spectral lines caused by a time delay or advance on the pulse propagation of the order of $\Delta t$. This way, the square of the geodesic separation can be approximated by
\begin{equation}
    2\sigma = (r + \Delta t)^2 - r^2 \approx 2r \Delta t.
\end{equation}
Hence, we can estimate the typical change in the photon's flight time when compared to the classical trajectory by \cite{ford1995}
\begin{equation}\label{timeShift_approx}
    \Delta t \approx \frac{\sqrt{|\langle\sigma_1^2\rangle_R|}}{r},
\end{equation}
where $\langle\sigma_1^2\rangle_R$ is the renormalized expectation value of $\sigma_1^2$, obtained by subtracting the unperturbed metric contribution. Note that $\langle\sigma_1^2\rangle$ is not necessarily positive since subvacuum effects may take place \cite{DeLorenci:2018moq, Wu:2008am}. An alternative derivation of Eq. \eqref{timeShift_approx} can be found in Ref. \cite{hongweiYu2009}. Note also that, since the graviton state defines a preferred frame of reference \cite{ford1995, Ellis1999}, the fact that the speed of light can be slightly higher or lower than $c$ it does not mean that causality is violated as the system is not naturally Lorentz invariant.
The calculation of $\langle\sigma_1^2\rangle$ can be extremely difficult depending on the starting point. However, since we are interested only in gravitational wave perturbations, that is, we are not specially interested in the gravitational waves source, we can adopt a quantization procedure that only retains freely-propagating degrees of freedom. Hence, we shall work with the transverse and tracefree (TT) gauge fixed by the conditions $\tensor{h}{_i^i} = \partial_j h^{ij} = h^{0\nu} = 0$. The procedure to obtain an expression for $\langle\sigma_1^2\rangle$ is discussed in details in Refs. \cite{ford1995, hongweiYu2009, lightconeCorda} and it reads 
\begin{equation}\label{sigma_1^2}
    \langle\sigma_1^2\rangle = \frac{1}{8} (r_1-r_0)^2 \int_{r_0}^{r_1} dr \int_{r_0}^{r_1} dr' \: n^i n^j n^k n^l G_{ijkl}(x,x'),
\end{equation}
where
\begin{equation}\label{G_ijkl}
    G_{ijkl}(x,x')=\langle h_{ij}(x)h_{kl}(x')+h_{ij}(x')h_{kl}(x)\rangle
\end{equation}
is the graviton two-point function, $dr = |d\textbf{x}|$ is the spatial interval of the trajectory of a null ray from a point $r_0$ to a point $r_1$ and $n^i = dx^i/dr$ is an unitary vector pointing to the direction of the geodesic. Direct calculations of the quantities in Eqs. \eqref{sigma_1^2} and \eqref{G_ijkl} will entail divergences. However, these divergences can be removed by subtracting the background spacetime respective contribution. The new renormalized result will provide us with a finite quantity when evaluated on the lightcone, and it will be zero if the quantum state of the gravitons is the unperturbed metric vacuum state, which in our case, is the Minkowski vacuum state.
It is worth mentioning that so far there was no need to particularize our equations to a set number of spacetime dimensions. All the expressions in this section are equally valid for any $(d+1)$-dimensional spacetime. Furthermore, $h_{ij}$ plays a significant role in obtaining a closed expression for $\langle\sigma_1^2\rangle$, as we can see from Eqs. \eqref{sigma_1^2} and \eqref{G_ijkl}. In the next section, we shall particularize our solutions to a five-dimensional spacetime and explicitly write $h_{\mu\nu}$ in order to compute $\langle\sigma_1^2\rangle_R$.
%
%
\section{Graviton two-point function in the five dimensional flat spacetime}
\label{sec:gvt_2ptFunc}
%
%
%
%
%
The line element describing our $(4+1)$-dimensional flat spacetime is given by \eqref{lineElement_5D} with $\mu, \nu =  0,1,2,3,4$. Additionally, in the TT gauge, Einstein's field equation for $h_{\mu\nu}$ reduces to a Klein-Gordon-like equation \cite{Misner1973, carroll2003spacetime}. This means that the perturbation can be written in terms of a plane wave expansion, that is
\begin{equation}\label{h_planWaveSol}
    h_{\mu\nu} = \sum_{\textbf{k},\lambda} \left[a_{\textbf{k},\lambda}e_{\mu\nu} (\textbf{k},\lambda) f_\textbf{k}(x) + \text{H.c.}\right],
\end{equation}
where \textbf{k} is the 4-dimensional wave vector in cartesian coordinates, $\lambda$ labels the polarization states, $e_{\mu\nu}(\textbf{k},\lambda)$ is the polarization tensor, $a_{\textbf{k},\lambda}$ and its Hermitian conjugate are the creation and annihilation operators associated with the behavior described in Eq. \eqref{h+h-}, and 
\begin{equation}\label{fk_5D_genSol}
    f_{\textbf{k}}(x) = A(2\omega)^{-\frac{1}{2}}e^{-i\omega t + i\textbf{k}\cdot \textbf{x}} = A(2\omega)^{-\frac{1}{2}}e^{-i\omega t}\phi_\textbf{k}(\textbf{x})
\end{equation}
and its Hermitian conjugate are solutions of the Klein-Gordon equation, with $A$ being a normalization constant. Substitution of Eq. \eqref{h_planWaveSol} in \eqref{G_ijkl}, while taking Eq. \eqref{h+h-} into consideration, yields
\begin{equation}\label{G_ijkl_planeWave}
    G_{ijkl}(x,x') = 2\text{Re} \sum_{\textbf{k},\lambda} e_{ij} (\textbf{k},\lambda)e_{kl} (\textbf{k},\lambda) f_\textbf{k}(x)f^{*}_\textbf{k}(x').
\end{equation}

Moreover, since $h_{\mu\nu}$ is purely spatial in the TT gauge, we can freely deal only with $h_{ij}$ instead. The sum over the polarization modes for a $(d+1)$-dimensional spacetime was obtained in Ref. \cite{hongweiYu2009}, and for our 5-dimensional spacetime is given by
\begin{equation}\begin{split}\label{polSum_general}
    \sum_\lambda e_{ij}(\textbf{k},\lambda)e_{kl}(\textbf{k},\lambda)  =\: &\delta_{ik}\delta_{jl} + \delta_{il}\delta_{jk} - \frac{2}{3}\delta_{ij}\delta_{kl} + \frac{4}{3}\hat{k}_i\hat{k}_j\hat{k}_k\hat{k}_l\\
    & + \frac{2}{3}\left(\hat{k}_i\hat{k}_j\delta_{kl}+\hat{k}_k\hat{k}_l\delta_{ij}\right) - \hat{k}_i\hat{k}_l\delta_{jk}-\hat{k}_i\hat{k}_k\delta_{jl}-\hat{k}_j\hat{k}_l\delta_{ik}-\hat{k}_j\hat{k}_k\delta_{il},
\end{split}\end{equation}
where $\hat{k}_i = k_i/|\textbf{k}|$ and $\delta_{ij}$ is the Kroeneker delta. Eq. \eqref{sigma_1^2} tells us that the only contributions of the graviton two-point function for $\langle\sigma_1^2\rangle$ will be the ones corresponding to the direction of the geodesic. If we particularize our results for a photon traveling along a specific direction, let us say $z$, then Eq. \eqref{polSum_general} reduces to
\begin{equation}
     \sum_\lambda e_{zz}(\textbf{k},\lambda)e_{zz}(\textbf{k},\lambda)  = \frac{4}{3} - \frac{8}{3}\hat{k}_z\hat{k}_z + \frac{4}{3}\hat{k}_z\hat{k}_z\hat{k}_z\hat{k}_z,
\end{equation}
and Eq. \eqref{G_ijkl_planeWave} becomes
\begin{equation}\begin{split}\label{Gzzzz_planeWaveExp}
    G_{zzzz}(x,x') & = \frac{8}{3}\text{Re} \sum_{\textbf{k}}\left(1 - 2 \frac{k_z^2}{|\textbf{k}|^2} +  \frac{k_z^4}{|\textbf{k}|^4}\right) f_\textbf{k}(x)f^{*}_\textbf{k}(x')\\
    & = \frac{8}{3}\left[D(x,x')-2F_{zz}(x,x')+H_{zzzz}(x,x')\right],
\end{split}\end{equation}
where $D(x,x')$ is the real part of the Wightman function for $f_\textbf{k}(x)$. Using Eq. \eqref{fk_5D_genSol} we can write $F_{zz}(x,x')$ and $H_{zzzz}(x,x')$ as
\begin{equation}\label{Fzz_planeWaveExp}
    F_{zz}(x,x') = 
    - \text{Re}\: \partial_{\Delta z}^2 \sum_\textbf{k} \frac{e^{-i\omega \Delta t}}{2\omega^3} \f_\textbf{k}(\textbf{x})\f^{*}_\textbf{k}(\textbf{x}'),
\end{equation}
and, 
\begin{equation}\label{Hzzzz_planeWaveExp}
    H_{zzzz}(x,x') = 
    \text{Re}\: \partial_{\Delta z}^4 \sum_\textbf{k} \frac{e^{-i\omega \Delta t}}{2\omega^5} \f_\textbf{k}(\textbf{x})\f^{*}_\textbf{k}(\textbf{x}').
\end{equation}
Up to this point we have revisited the lightcone fluctuations formalism and obtained an expression for the graviton two-point function in five dimensions adopting the TT gauge. In the next section we will turn our attention to understanding how the quantization of $h_{\mu\nu}$ and the imposition of a quasiperiodic condition to the extra dimension affect the flight time of a photon propagating along an uncompactified one. This could provide us with an interesting test to the existence of extra dimensions since it would depend upon photon detection.

\section{Lightcone fluctuations in a Kaluza-Klein model for a photon propagating in an ordinary dimension}
\label{sec:LC_KK}
%
%
\subsection{Lightcone fluctuation for a photon propagating along the $z$-direction} 
\label{subsec:sub_LCalongZ}
%
In the past century, following the formulation of General Relativity, several attempts have been made to unify the electromagnetic theory with gravity \cite{kaluzaOriginal,kleinOriginal,weyl1918, einstein1931}. In particular, in 1921, Theodor Kaluza proposed a five-dimensional world in an attempt to incorporate the electromagnetic tensor into the metric tensor \cite{kaluzaOriginal} and five years later, Oskar Klein connected Kaluza's approach to the emergent quantum mechanics \cite{kleinOriginal}. Kaluza's work is regarded to be the first attempt to grasp a world beyond our usual $(3+1)$ framework. Consequently, scenarios involving $(4+1)$ dimensions featuring a small unobservable extra dimension with quantized fields are often regarded as the Kaluza-Klein original model.

\begin{figure}[h]
    \includegraphics[scale=0.25]{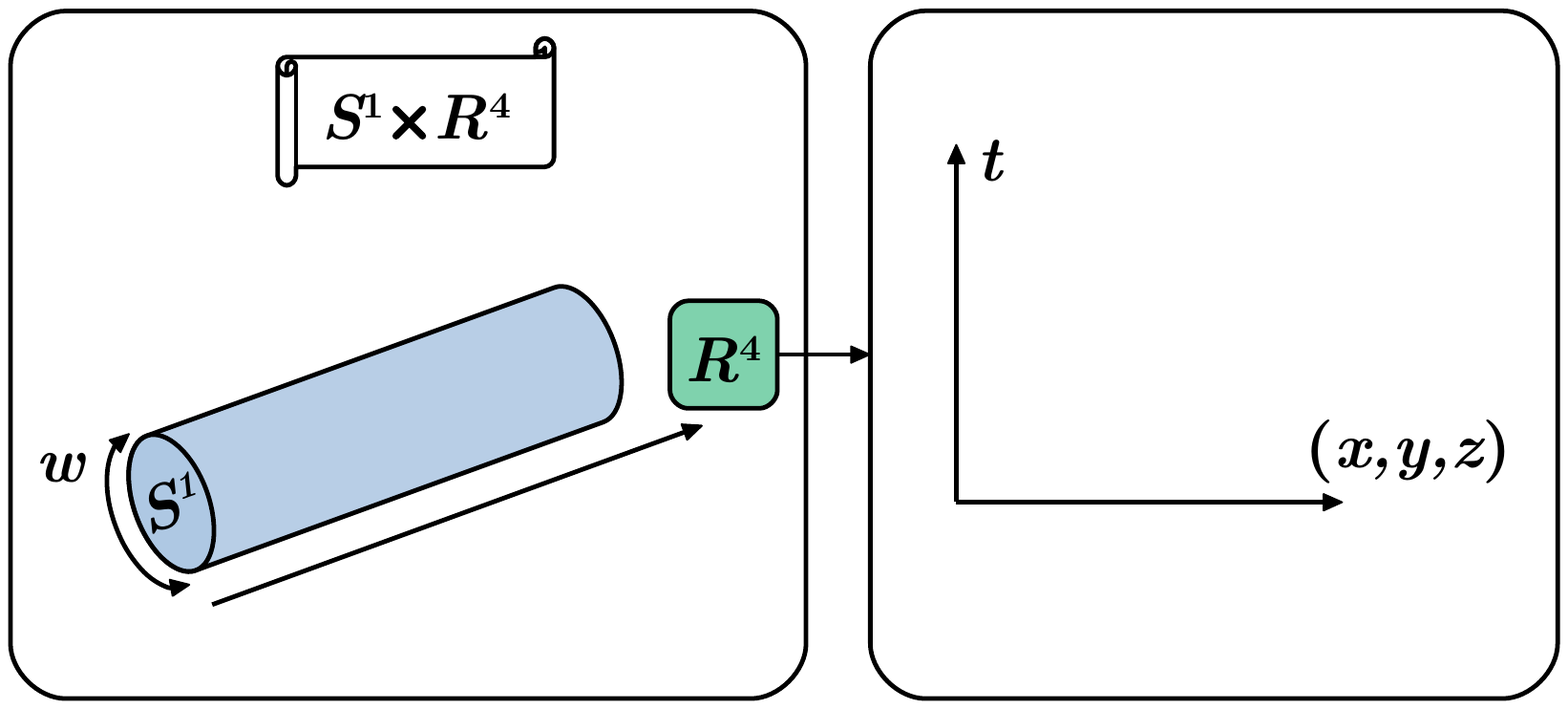}
     \includegraphics[scale=0.25]{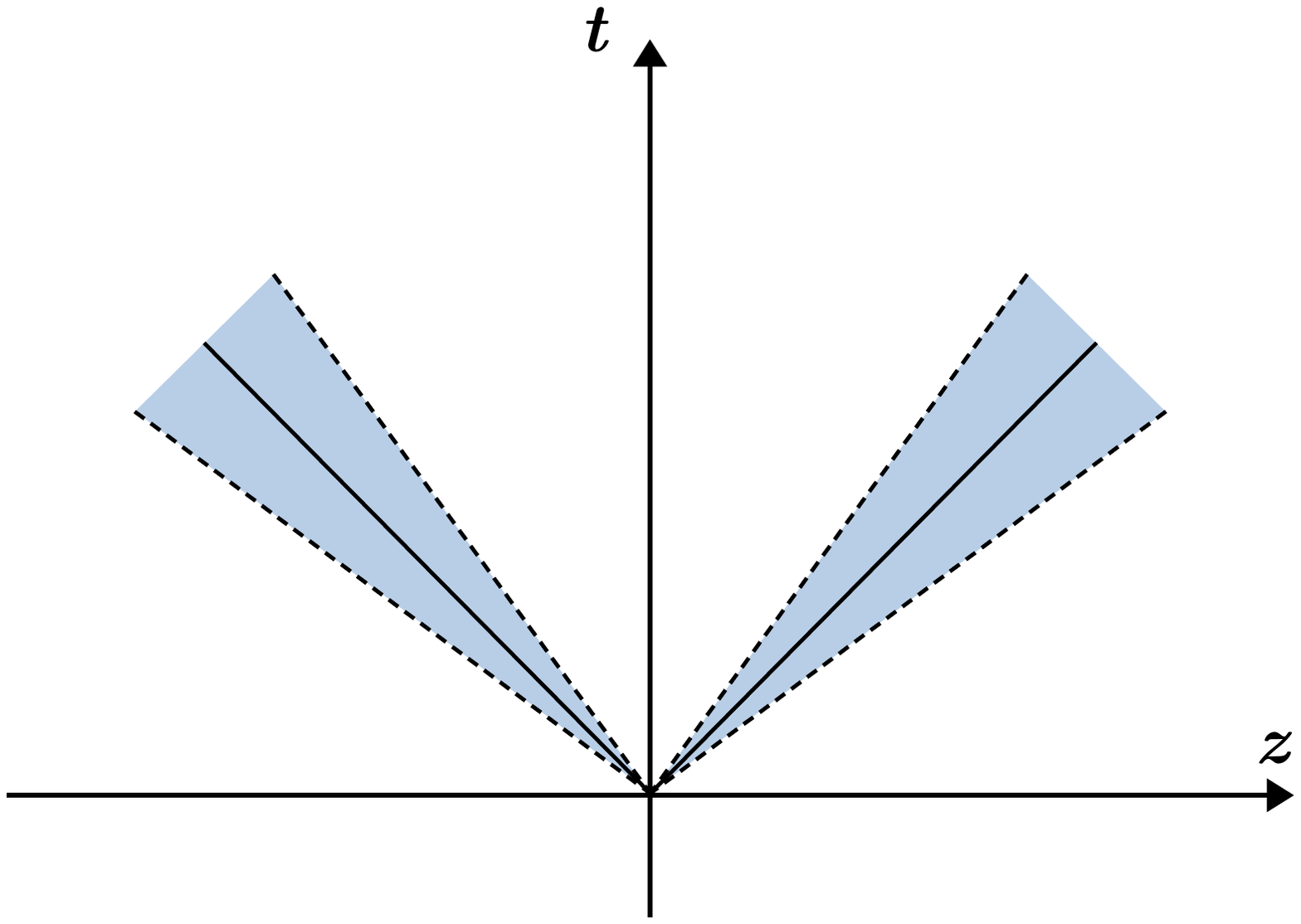}
    \caption{Illustrations of the compactified extra dimension (left) and of a photon propagating in the $z$-direction (right). The latter is based on illustration from Ref. \cite{ford1995} and shows the fluctuations around the fixed lightcone (solid black line).}
    \label{extra_light}
\end{figure}

In this work we compactify the extra dimension by imposing a quasiperiodic condition (see Fig.\ref{extra_light}), which reads
\begin{equation}\label{quasiP_BC}
    \f_\textbf{k}(x,y,z,w+\ell_c) = e^{i2\pi\alpha}\f_\textbf{k}(x,y,z,w),
\end{equation}
where $\phi_\textbf{k}(\textbf{x})$ is the spatial part of the quantized perturbation given by Eqs. \eqref{h_planWaveSol} and \eqref{fk_5D_genSol}, $\ell_c$ is the compactification length and $\alpha$ is a parameter varying in the range, $0 \leq\alpha < 1$, which regulates the phase angle $2\pi\alpha$. Note that Eq. \eqref{quasiP_BC} reduces to the periodic case discussed in Ref. \cite{hongweiYu2009} for $\alpha = 0$, whereas $\alpha = 1/2$ corresponds to the antiperiodic case. 

The imposition of \eqref{quasiP_BC} to the free solution of the five-dimensional Klein-Gordon equation for the massless case in flat spacetime will cause the discretization of the momentum coordinate parallel to the direction of compactification. Hence, we have
\begin{equation}\label{kw_5D}
    k_w = \frac{2\pi}{\ell_c}(n+\alpha), \quad\qquad\qquad n = 0, \pm 1, \pm 2, \dots\;.
\end{equation}
Thus, the complete set of normalized spatial solutions under the quasiperiodic condition takes the form 
\begin{equation}\label{phi_5D}
    \f_\textbf{k}(x) = \frac{e^{i\textbf{k}_T \cdot\textbf{r}_T + i \frac{2\pi (n+\alpha)}{\ell_c}w}}{\sqrt{(2\pi)^3 \ell_c}},
\end{equation}
where $\textbf{k}_T$ and $\textbf{r}_T$ are the momenta and the corresponding spatial coordinate 3-vectors relative to the directions perpendicular to the extra dimension. The eigenvalue equation reads
\begin{equation}\label{eigValueEq_5D}
    \omega^2 = |\textbf{k}|^2 = k_x^2 + k_y^2 + k_z^2 + \left(\frac{2\pi}{\ell_c}\right)^2\left(n+\alpha\right)^2.
\end{equation}
The sum over all momenta coordinate possible values in Eqs. \eqref{Gzzzz_planeWaveExp}, \eqref{Fzz_planeWaveExp}, and \eqref{Hzzzz_planeWaveExp} in this scenario should be written as
\begin{equation}\label{momentaSum_5D}
    \sum_\textbf{k} = \int_{-\infty}^{\infty} dk_x\int_{-\infty}^{\infty} dk_y\int_{-\infty}^{\infty} dk_z \sum_{n=-\infty}^\infty.
\end{equation}
Our objective now is to compute the renormalized expectation value of $\sigma_1^2$ for a photon that propagates in a direction perpendicular to the compactification. Thus, let us assume that the photon is propagating along the $z$-axis direction from a point $a$ to a point $b$, see Fig.\ref{extra_light} (right). Thus, Eq. \eqref{sigma_1^2} becomes
\begin{equation}\label{sigma_1^2_Z}
    \langle\sigma_1^2\rangle_R = \frac{1}{8}(b-a)^2 \int_{a}^{b} dz \int_{a}^{b} dz' G^R_{zzzz}(t,z,t',z')|_{\Delta t = \Delta z},
\end{equation}
where $G^R_{zzzz}(t,z,t',z')$ is given by Eq. \eqref{Gzzzz_planeWaveExp} with the renormalized functions $D^R(x,x')$, $F^R_{zz}(t,z,t',z')$ and $H^R_{zzzz}(t,z,t',z')$ calculated in the appendixes. By substituting Eqs. \eqref{DR_5D_photonInZ}, \eqref{Fzz_5D_QP_R} and \eqref{Hzzzz_5D_QP_R} in Eq. \eqref{Gzzzz_planeWaveExp}, performing the derivatives and then taking the null interval $\Delta z = \Delta t$, we obtain the renormalized graviton two-point function evaluated on the lightcone
\begin{equation}\label{Gzzzz_R_final}
    G_{zzzz}^R(t,z,t',z')|_{\Delta t = \Delta z} = \frac{16}{3\pi^2} \sum_{n=1}^\infty \frac{\Delta z\left(n\ell_c\right)\left(5n^2 \ell_c^2-3\Delta z^2\right)}{(\Delta z^2+n^2 \ell_c^2)^5} \cos (2\pi n \alpha).
\end{equation}
Finally, by substituting Eq. \eqref{Gzzzz_R_final} in Eq. \eqref{sigma_1^2_Z} and performing the integrals, we find
\begin{equation}\label{sigma_1^2_final_Z}
    \langle\sigma_1^2\rangle_R = \frac{2\ell_c}{9\pi^2}\sum_{n=1}^\infty \frac{\gamma^8\cos (2\pi n \alpha) }{n(n^2+\gamma^2)^3},
\end{equation}
where $r=b-a$ and $\gamma = r/\ell_c$. The exact sum in Eq. \eqref{sigma_1^2_final_Z} for any $\alpha$ can in fact be obtained, however, the resulting expression is too lengthy to be exposed here. Alternatively, we shall analyze some cases of interest in parallel to the exact behaviour of Eq. \eqref{sigma_1^2_final_Z}. The search for the existence of extra-dimensions permeates many phenomena in physics. In particular, direct measurements of the gravitational force between two masses have shown that the inverse-square law, which is expected to be modified as the distance between the masses approaches the size of the extra dimension, holds down to $5.2\cdot 10^{-3}$cm \cite{Adelberger2007,Adelberger2020}. On the other hand, the smallest distance ever probed by the Large Hadron Collider (LHC) without any indication of the existence of extra dimensions has been $10^{-18}$cm \cite{ATLAS:2012hvw}. As this distance refers to the energy scale in which experiments are conducted in the LHC, it should be understood as a measure of how close such experiments are to the Planck scale. For instance, as the compactification length can be related to the fundamental Planck length in five dimensions $\ell_P^{(5)}$, as well as to the known four-dimensional Planck length $\ell_P$, through the expression \cite{zwiebach_2004}
\begin{equation}\label{Lplanck_in4and5_Lcompac}
    \ell_P^{(5)} = \left(\ell_c \ell_P^2\right)^\frac{1}{3},
\end{equation}
the exact value of $\ell_P^{(5)}$ depends on how gravity works in a five-dimensional world and, as a consequence, there is no guarantee it will be equal the four-dimensional Planck length $\ell_P$. If experiments on the LHC were close to the fundamental Planck scale, i.e., if we were to have $\ell_P^{(5)}\sim 10^{-19}$cm, Eq. \eqref{Lplanck_in4and5_Lcompac} would tell us that the length of the extra dimension would be around ten times the distance from the Earth to the Moon. Such large extra dimension would have been already detected and, therefore, if we do live in a $5$-dimensional world, the energy scale at which the LHC operates at the moment is not enough to probe it. Note that one of the most significant indicators of the existence of extra dimensions through the LHC would be the creation of heavier particles \cite{Agashe:2020wph} whereas, for the case we are discussing, this indicator would be the deviation on the typical flight time of a photon, detected by instruments such as spectrographs. Such equipments use a combination of mirrors and filters to divide the light in several wavelength intervals. Since there exists an optical path inside the detector, the distance $r$ traveled by the photon would be much larger than the smallest distance for which the inverse-square law holds, consequently, the cases of most interest for us will be those where $\gamma\gg 1$.


%
%
\subsection{The periodic case ($\alpha = 0$) and the Planck length in 5D}
\label{subsec:alphaNull}
%
%
The particular case $\alpha = 0$ in Eq. \eqref{sigma_1^2_final_Z} is exactly the expectation value of $\sigma_1^2$ found in Ref. \cite{hongweiYu2009}. However, let us revisite it here by performing the sum over $n$ for $\alpha=0$, that is,
\begin{equation}\begin{split}\label{sigma1^2_alphaNull_exact}
    \langle\sigma_1^2\rangle_R = &  \frac{\ell_c\gamma^2}{72\pi^2}  \Big\{16 \gamma_e + 8\left[\psi^{(0)}(1 - i \gamma) + \psi^{(0)}(1 + i \gamma)\right] \\
    & \quad\quad\quad  + 5 i\gamma \left[\psi^{(1)}(1 - i \gamma) - \psi^{(1)}(1 + i \gamma)\right] - \gamma^2 \left[\psi^{(2)}(1 - i \gamma) + \psi^{(2)}(1 + i \gamma)\right]\Big\},
\end{split}\end{equation}
where $\gamma_e$ is the Euler-Mascheroni constant and $\psi^{(n)}(x)$ is the Polygamma function. This expression is sufficient for us to compute the exact mean deviation on the photon flight time in Eq. \eqref{timeShift_approx}. In order to have a better understanding of what these results are indicating, we shall expand \eqref{sigma1^2_alphaNull_exact} for $\gamma \gg 1$, i.e.,
\begin{equation}\label{sigma1^2_alphaNull_largeGamma}
    \langle\sigma_1^2\rangle_R = \frac{2\ell_c\gamma^2}{9\pi^2}\left[\left(\gamma_e - \frac{3}{4}+\text{ln}\gamma\right) + \frac{1}{4\gamma^2} + \mathcal{O}(\gamma^{-4})\right].
\end{equation}
By combining Eqs. \eqref{sigma1^2_alphaNull_largeGamma} and \eqref{timeShift_approx} we can write
\begin{equation}\label{DeltaT_alphaNull_largeGamma}
    \Delta t \approx \sqrt{\frac{2}{9\pi^2 \ell_c} \text{ln}\gamma} = \ell_c^{-\frac{1}{2}}\sqrt{\frac{2\text{ln}\gamma}{9\pi^2}}.
\end{equation}
We should note that, even if $r$ is a cosmological distance and $\ell_c$ is near the four-dimensional Planck scale, the term inside the square root symbol in Eq. \eqref{DeltaT_alphaNull_largeGamma} is still not large enough to have a significant impact on the magnitude of $\Delta t$, as it has been pointed out in Ref. \cite{hongweiYu2009}. However, note also that as the compactification length $\ell_c$ decreases the time shift becomes greater, something that can make the result in Eq. \eqref{DeltaT_alphaNull_largeGamma} detectable. 
%

By recovering units of time in Eq. \eqref{DeltaT_alphaNull_largeGamma} we find that, for $\alpha=0$ and $\gamma\gg 1$,
\begin{equation}\label{DeltaT_approx_plancktime}
    \Delta t \approx \ell_c^{-\frac{1}{2}}\frac{1}{c} \left(\frac{G\hbar}{c^3}\right)^\frac{3}{4}\sqrt{\frac{2\text{ln}\gamma}{9\pi^2}} = t_P \left(\frac{\ell_P}{\ell_P^{(5)}}\right)^\frac{3}{2}\sqrt{\frac{2\text{ln}\gamma}{9\pi^2}},
\end{equation}
where $t_P=\ell_P/c$ is the four-dimensional Planck time and Eq. \eqref{Lplanck_in4and5_Lcompac} has been used, with $\ell_P=\left(\frac{\hbar G}{c^3}\right)^{\frac{1}{2}}$. From Eq. \eqref{DeltaT_approx_plancktime} we can see that, if the compactification length is close to the four-dimensional Planck scale, or equivalently, from Eq. \eqref{Lplanck_in4and5_Lcompac}, if $\ell_P^{(5)}\sim\ell_P$, the mean flight time deviation will be of order of the four-dimensional Planck time $t_P \sim 10^{-44}$s. Such deviation would be far from observable. This is in essence the same conclusion reached by the authors in Ref. \cite{hongweiYu2009}. 

\begin{figure}[h]
    \includegraphics[scale=1.2]{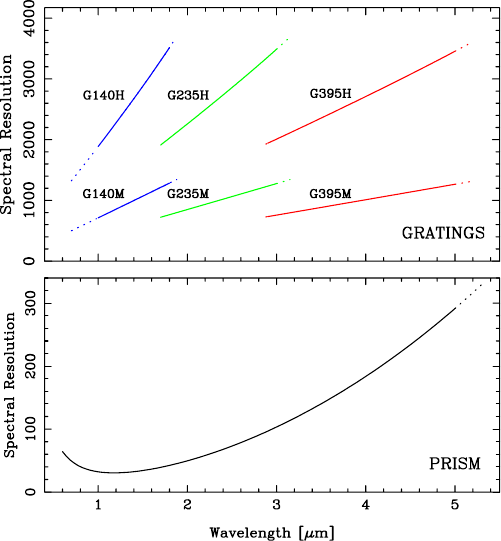}
    \caption{Reproduced without modifications from Jakobsen, P., Ferruit, et al., \textit{The Near-Infrared Spectrograph (NIRSpec) on the James Webb Space Telescope} (JWST) published in A\&A by EDP Sciences (Ref. \cite{ResolvingPowerNIRSPec2022}) under the CC BY 4.0 \href{https://creativecommons.org/licenses/by/4.0/legalcode.en}{Legal Code}. Depiction of the spectral resolution for observed wavelength for each of the seven dispersers on the GWA from the NIRSpec, being one prism and six gratings. In order to obtain an detection throughout the NIRSpec full wavelength range, one could rely on a single exposure using the prism. The result would be a low resolution and highly contaminated image. In order to obtain high resolution data with least contamination containing the full wavelength range for which the NIRSpec was designed, each of the six gratings on the GWA work combined only with specific transmission filter from the seven present in the filter wheel assembly (FWA). Each solid line depicts the nominal spectral resolution (the resolution at the center of the nominal wavelength range for a given disperser-filter combination) for each disperser-filter combination (further details on filter-disperser pairing can be found in the original source and in the JWST documentation), and the dotted lines depicts the anticipated wavelength range.}  
    \label{fig:NIRSpec_specs}
\end{figure}

However, from another perspective, let us see possible estimations  by considering a measure device based on a spectrograph. When a spectrograph or radio telescope observes light emitted from a source at certain wavelength $\lambda$ with a resolving power $R$, it will be able to discern shifts in wavelength in the order of $\Delta\lambda = \lambda/R$. For instance, the NIRSpec aboard the James Webb Space Telescope is able to detect photons in a wavelength range of $6\times 10^{-5}$cm to $5.3\times 10^{-4}$cm, with spectral resolutions of 100, 1000, and 2700 \cite{ResolvingPowerNIRSPec2022}. More specific values of the spectral resolution for each wavelength observed from the NIRSpec are depicted in Fig.\ref{fig:NIRSpec_specs}, reproduced without modifications from Ref. \cite{ResolvingPowerNIRSPec2022}. The Figure shows performances for each of the seven dispersive elements present in the grating wheel assembly (GWA) on the NIRSpec, one prism and 6 gratings. It is clear from the plot that, even though it covers all the wavelength range for which the NIRSpec was designed, the spectral resolution curve for the prism does not provide the optimal $\Delta\lambda$. For this end one should look at the G140H data. In order to detect deviations on the flight time, it is more relevant to look for the smallest observed wavelength to resolving power ratio of the detector rather than the distance from the source. Thus, by supposing that a detection occurs at $\lambda = 1.4\times 10^{-4}$cm with resolving power $R=2700$, the perceived shift in wavelength would be of order of $\Delta \lambda \approx 5.2\times10^{-8}$cm. In this sense, the smallest discernible flight time shift for the NIRSpec in these conditions would be $\Delta t\approx 1.7\times 10^{-18}$s. Substituting this flight time shift back in Eq. \eqref{DeltaT_approx_plancktime} we find that, in order to obtain an observable flight time deviation, the extra dimension length should be of order of $\ell_c\sim 10^{-84}$cm. Equivalently, the fundamental Planck length in five dimensions should be no greater than $\ell_P^{(5)} \sim 10^{-50}$cm, many orders of magnitude smaller than $\ell_P\simeq 10^{-33}$cm.
The calculations of $\Delta t$ become even more difficult when considering a Kaluza-Klein model with higher than five dimensions. However, if we were to add more extra dimensions, we expect that for each extra compact dimension considered, $\Delta t$ would depend on $\ell_c$ with bigger negative powers, which can cause such deviations to become more significant even for extra dimensions with greater length than our current estimative. 

Now we could also wonder whether the metric fluctuation effects deriving from different values of $\alpha$ in our model could be stronger than the periodic case. The periodic case $\alpha=0$ in fact maximizes the effects of the lightcone fluctuations on the flight time mean deviation of photons, as we can see from the cosine in Eq. \eqref{sigma_1^2_final_Z}. In the next subsection we will find a more compact expression for other values of the phase regulator $\alpha$.
%
%
\subsection{Other values of $\alpha$}
\label{subsec:alphaNonNull}
%
%
Since the most interesting cases for us involve $\gamma\gg 1$, let us find  a more convenient and compact expression than Eq. \eqref{sigma_1^2_final_Z} by writing
\begin{equation}\label{sigma1^2_nbar}
    \langle\sigma_1^2\rangle_R = \frac{2\ell_c\gamma^2}{9\pi^2}\sum_{n=1}^\infty \frac{\cos (2\pi n\alpha)}{n(\Bar{n}^2+1)^3},
\end{equation}
where $\Bar{n} = n/\gamma$. Note that the sum in Eq. \eqref{sigma1^2_nbar} is a descending series of $n$ and, large values of $\gamma$ requires that $\Bar{n}<1$ in the dominating terms of the series. Using the negative binomial expansion
\begin{equation}
    (x+1)^{-n} = \sum_{k=0}^\infty \binom{n+k-1}{k}(-x)^k,
\end{equation}
which is convergent for $|x|<1$, and
\begin{equation}
    \binom{n}{k} = \left\{
    \begin{matrix}
        \frac{n!}{k!(n-k)!}, \quad 0 \leq k < n\\
        0, \quad \text{else}
    \end{matrix}
    \right.,
\end{equation}
we are able to write \eqref{sigma1^2_nbar} as
\begin{equation}
    \langle\sigma_1^2\rangle_R = \frac{2\ell_c\gamma^2}{9\pi^2} \sum_{k=0}^\infty (-1)^k\frac{(k+2)(k+1)}{2\gamma^{2k}}\sum_{n=1}^\infty n^{2k-1}\cos(2\pi n \alpha).
\end{equation}
The sum in $n$ have now taken the form of Polylogarithm functions, namely 
\begin{equation}
    \langle\sigma_1^2\rangle_R = \frac{2\ell_c\gamma^2}{9\pi^2} \sum_{k=0}^\infty (-1)^k\frac{(k+2)(k+1)}{4\gamma^{2k}} \left[\text{Li}_{1-2k}(e^{i2\pi\alpha})+\text{Li}_{1-2k}(e^{-i2\pi\alpha})\right].
    \label{approx}
\end{equation}
It is clear from the remaining sum over $k$ that this is indeed an expression for large $\gamma$. Evaluating the Polylogarithm functions for the first two values of $k$ allows us to write
\begin{figure}[h]
    \includegraphics[scale=0.6]{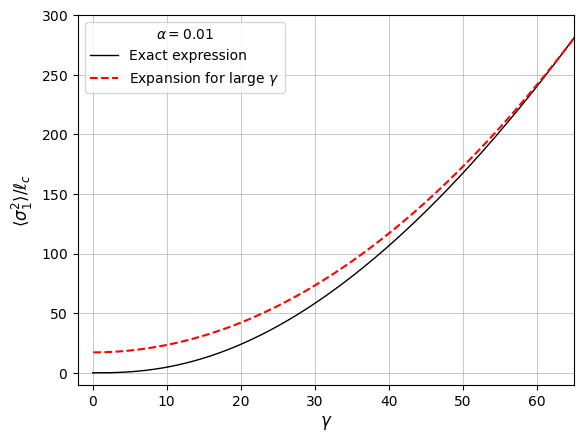}
    \caption{In terms of $\gamma$, the behaviour for $\alpha=0.01$ of the exact sum on Eq. $\eqref{sigma_1^2_final_Z}$, in units of $\ell_c$, is depicted in solid black line and compared with the expansion for $\gamma\gg 1$ in Eqs. \eqref{sigma1^2_alphaNull_largeGamma} and \eqref{sigma1^2_alphaNonNull_largeGamma} in dotted red line. In fact, for large $\gamma$, the curves meet for any value of $\alpha$.}
    \label{fig:sigma}
\end{figure}
\begin{equation}\label{sigma1^2_alphaNonNull_largeGamma}
    \langle\sigma_1^2\rangle_R = \frac{\ell_c\gamma^2}{18\pi^2}\left[-2\ln\left(1-e^{-i2\pi\alpha}\right)-2\ln\left(1-e^{i2\pi\alpha}\right)+\frac{3\csc^2(\pi\alpha)}{\gamma^2} + \mathcal{O}(\gamma^{-4})\right],
\end{equation}
which is valid for $\alpha\neq 0$. For $\alpha = 0$ and large $\gamma$ one should use Eq. \eqref{sigma1^2_alphaNull_largeGamma}. Furthermore, it also becomes clear from Eq. \eqref{sigma1^2_alphaNonNull_largeGamma} that there are certain values of $\alpha$ for which $\langle\sigma_1^2\rangle_R< 0$, i.e., subvacuum effects are present in our system \cite{DeLorenci:2018moq, Wu:2008am}. Additionally, one can see directly from Eq. \eqref{sigma_1^2_final_Z} that, among all $\alpha$ resulting in negative $\langle\sigma_1^2\rangle_R$, the antiperiodic case will result in the highest $|\langle\sigma_1^2\rangle_R|$, and therefore, in the highest flight time mean deviation. The behavior of the exact expression for $\langle\sigma_1^2\rangle_R$ in Eq. \eqref{sigma_1^2_final_Z}, as well as the expansion for large $\gamma$ in Eq. \eqref{sigma1^2_alphaNonNull_largeGamma} after discarding terms of $k\geq 2$ are depicted in Fig.\ref{fig:sigma} for $\alpha = 0.01$ in solid black and dotted red lines, respectively. It becomes clear from the plot that Eq. \eqref{sigma1^2_alphaNonNull_largeGamma} is in fact a good expression for larger values of $\gamma$. Note that, although we have taken a particular value of $\alpha$ to show in Fig.\ref{fig:sigma}, we have numerically verified that the approximation \eqref{approx} is good for any value of $\alpha\neq 0$.
Note also that in the plot of Fig.\ref{fig:sigma} both $\gamma=r/\ell_c$ and $\langle\sigma_1^2\rangle/\ell_c$, are dimensionless quantities, as we can see from any of the previous equations for $\langle\sigma_1^2\rangle$. This choice was made simply to enhance clarity. 

Note that Eq. \eqref{sigma1^2_alphaNonNull_largeGamma} contains a dependence on $\gamma$ with positive and negative powers, in contrast with the expression \eqref{sigma1^2_alphaNull_largeGamma} for $\alpha = 0$ that contains also a logarithmic dependence. Besides, Eq. \eqref{sigma1^2_alphaNonNull_largeGamma} also depends on $\alpha$. This will result in much larger  values of $\langle\sigma_1^2\rangle_R$ when $\alpha = 0$ than when any other value of the quasiperiodic parameter is assumed. By discarding contributions of the order of $\gamma^{-2}$ and lower in Eq. \eqref{sigma1^2_alphaNonNull_largeGamma} and substituting it back in Eq. \eqref{timeShift_approx}, one can also obtain an expression for the mean flight time deviation of the photons for $\alpha\neq 0$. Recovering units of time on the resulting expression yields
\begin{equation}\label{DeltaT_alphaNonNull_largeGamma}
    \Delta t \approx \frac{\ell_c^{-\frac{1}{2}}}{\sqrt{18\pi^2}}\frac{1}{c} \left(\frac{G\hbar}{c^3}\right)^\frac{3}{4}
    \Bigg|-2\ln\left(1-e^{-i2\pi\alpha}\right)-2\ln\left(1-e^{i2\pi\alpha}\right)+\frac{3\csc^2(\pi\alpha)}{\gamma^2}\Bigg|^{\frac{1}{2}}.
\end{equation}
We can see once again that the highest dependence on $\gamma$ for $\alpha\neq 0$ comes from the last term in the square root in the r.h.s. of Eq. \eqref{DeltaT_alphaNonNull_largeGamma}, in contrast to the square root of the logarithm obtained for $\alpha = 0$ in Eq. \eqref{DeltaT_alphaNull_largeGamma}. Consequently, for large values of $\gamma$, the dominating terms in the above expression will be the logarithmic ones fixed by the quasiperiodicity phase angle regulator $\alpha$. Hence, the main quantity guiding the magnitude of $\Delta t$, for a fixed $\alpha$, will once again be the inverse of the square root of the compactification length, namely, $\ell_c^{-1/2}$. 
\begin{figure}[h]
    \includegraphics[scale=0.60]{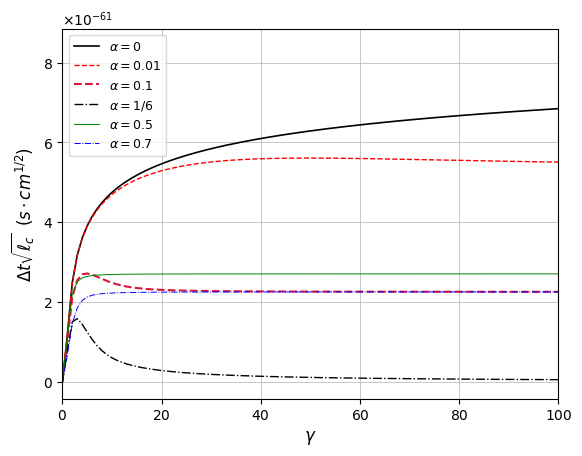}
    \caption{The exact behaviour of $\Delta t\sqrt{\ell_c}$, in terms of $\gamma$, for several values of $\alpha$.}
    \label{fig:DeltaT}
\end{figure}
The behavior of the exact expression for $\Delta t$ obtained by combining Eqs. \eqref{sigma_1^2_final_Z} and \eqref{timeShift_approx} is depicted in Fig.\ref{fig:DeltaT}. Note that the quantity on the vertical axis, $\Delta t\sqrt{\ell_c}$, has units of $s\cdot cm^{1/2}$. This choice enables a direct assessment of the influence of both $\gamma$ and $\ell_c$ on the magnitude of $\Delta t$ based on the observational criteria discussed in Sec. \ref{subsec:alphaNull} for the case $\alpha = 0$. For example, by looking at Fig.\ref{fig:DeltaT} one can see that, for large values of $\gamma$ (of order of $10^{2}$), one obtains $\Delta t\sqrt{\ell_c} \sim 10^{-61}\text{s}\cdot\text{cm}^{1/2}$. It follows from a straightforward dimensional analysis that, for an observable flight time shift ($\Delta t \sim 10^{-18}$s) one must have $\ell_c\sim 10^{-84}$cm, in perfect accordance with the discussion in the previous subsection. In our analysis, $\alpha$ can not be determined from a single measurement and, therefore, is a free parameter.
Fig.\ref{fig:DeltaT} also reflects the different behaviours of Eqs. \eqref{DeltaT_alphaNull_largeGamma} and \eqref{DeltaT_alphaNonNull_largeGamma} and, as a result, as $\gamma$ increases, the periodic case will lead to a logarithmically increasing mean flight time deviation, whereas for each other $\alpha\neq 0$ case will rapidly converge to a constant $\Delta t$. This distinction can be crucial in comprehending the potential structure of the extra dimension through measurements at a given distance $r$ from the photon source. Although this distance is limited to the length of the observable universe, approximately $10^{28}$ cm, the functional dependence of $\Delta t$ with $\gamma$ could play a pivotal role in determining values for $\alpha$ and, consequently, for $\ell_c$. 
%
\begin{figure}[h]
    \includegraphics[scale=0.68]{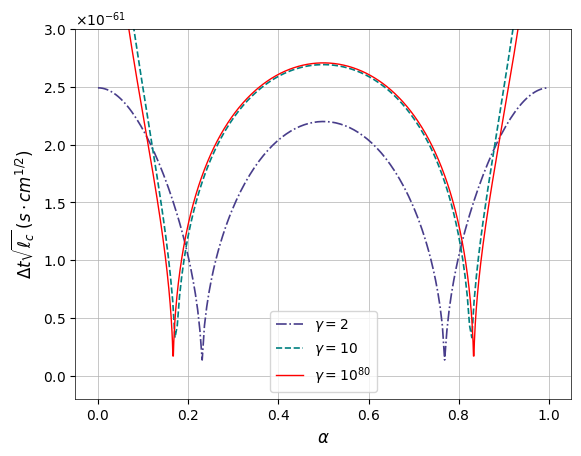}
    \caption{The behaviour of $\Delta t\sqrt{\ell_c}$ as a function of the quasiperiodic parameter $\alpha$, for some values of $\gamma$.}
    \label{fig:deltaT_Alpha}
\end{figure}

%

From another perspective, Fig.\ref{fig:deltaT_Alpha} depicts the general behaviour of $\Delta t\sqrt{\ell_c}$ as a function of $\alpha$, assuming three distinct values of $\gamma$. For each curve, the region in between the minima provides the values of $\alpha$ for which $\langle\sigma_1^2\rangle_R<0$, whereas the values of $\alpha$  outside this region give $\langle\sigma_1^2\rangle_R>0$. This can be verified from the exact expression in Eq. \eqref{sigma_1^2_final_Z}. As we have previously stated, it also becomes clear from the plot in Fig.\ref{fig:deltaT_Alpha} that, for a fixed $\gamma$, $\alpha=1/2$ indeed results in a maximum flight time mean deviation when compared with all other values of $\alpha$ for which $\langle\sigma_1^2\rangle_R < 0$. An interesting aspect also revealed by the plot is that there are more than one possible value for $\alpha$ that provides the same potential `measured' $\Delta t$, making it difficult to completely understand the structure of the extra dimension. In addition, we can also see in Fig.\ref{fig:deltaT_Alpha} the convergence of $\Delta t\sqrt{\ell_c}$ to constant values shown in the plot of Fig.\ref{fig:DeltaT} for $\alpha\neq 0$. Note in Fig.\ref{fig:deltaT_Alpha} that there are no difference in the curves for sufficiently large $\gamma$.
%

%
Upon the possibility that we live in a five-dimensional world, both the size of the extra dimension, $\ell_c$, and the phase angle regulator, $\alpha$, are crucial parameters to be inferred from experimental data, like the ones from NIRSpec. As we have mentioned before, the extra dimension should be small. If it exists, it is expected to be smaller than $5.2\cdot 10^{-3}$cm, the smallest distance for which it was verified that the gravitational inverse-square law holds. On the other hand, there are some factors that can modify the spectrum of a light pulse, if we assume that lightcone fluctuations are one of them and that the resulting mean flight time deviation falls right within the NIRSpec's sensitivity, then our model provides estimations on the size of the extra dimension for $\alpha=0$, and also for $\alpha\neq 0$, that are of order of $\ell_c\sim 10^{-84}$cm, a length much smaller than the sensitivity range of tabletop experiments and of the LHC. This also imposes an estimation on the Planck length in five dimensions that is of order of $10^{-50}$cm.
\section{Summary and discussion}
\label{conc}
%
We have studied the effects of lightcone fluctuations in a five dimensional spacetime with a quasiperiodically compactified extra dimension on a photon travelling along the $z$-axis. In this context, we found an expression for the linear perturbation $h_{\mu\nu}$ in Eq. \eqref{h_planWaveSol} combined with Eqs. \eqref{fk_5D_genSol} and \eqref{phi_5D}. In Eq. \eqref{Gzzzz_R_final} we found the renormalized graviton two-point function in flat 5D with the quasiperiodically compactified extra dimension and in Eq. \eqref{sigma_1^2_final_Z} we used it to compute the exact expression for the renormalized expectation value for the squared geodesic separation between two points $x$ and $x'$, namely $\langle\sigma_1^2\rangle_R$, which will lead to the mean deviation in Eq. \eqref{timeShift_approx} for a photon flight time. It is possible to perform the exact sum in Eq. \eqref{sigma_1^2_final_Z}, however, the resulting expression is too lengthy to be written down. This has not provided any trouble for our analysis since the cases of interest are the ones for which $\gamma=r/\ell_c \gg 1$. 

Consequently, we found an expression for large values of $\gamma$ for $\alpha =0$ in Eq. \eqref{sigma1^2_alphaNull_largeGamma} and discussed the corresponding mean flight time deviation in Eqs. \eqref{DeltaT_alphaNull_largeGamma} and \eqref{DeltaT_approx_plancktime} and also some observational aspects of these results, based on the NIRSpec's sensitivity. Hence, we have found that the change in the flight time for a periodically compactified extra dimension would only be detectable if the size of the extra dimension is of order of $10^{-84}$cm, which from Eq. \eqref{Lplanck_in4and5_Lcompac} would reflect in a five dimensional Planck length of approximately $10^{-50}$cm, several order of magnitude smaller than the Planck length in our usual $(3+1)$ dimensions. 

We have also obtained the expression for large $\gamma$ and $\alpha\neq 0$ in Eq. \eqref{sigma1^2_alphaNonNull_largeGamma} and plotted the behaviour of $\langle\sigma_1^2\rangle_R$, showing that Eqs. \eqref{sigma1^2_alphaNull_largeGamma} and \eqref{sigma1^2_alphaNonNull_largeGamma} provide a good approximation for Eq. \eqref{sigma_1^2_final_Z} for large values of $\gamma$. We also plotted the exact behaviour of $\Delta t$ for some values of $\alpha$ in Fig.\ref{fig:DeltaT}. We have discussed the plots and expressions, and pointed out that $\Delta t$ for $\alpha = 0$ logarithmically increases with $\gamma$, whereas for $\alpha\neq 0$ it rapidly converges to constant values. Furthermore, among all values of $\alpha$ resulting in $\langle\sigma_1^2\rangle_R<0$, the antiperiodic case results in the highest mean flight time deviation. We have plotted in Fig.\ref{fig:deltaT_Alpha} the behaviour of $\Delta t$ in terms of $\alpha$ for three different values of $\gamma$, in order to further illustrate some features of our results and discuss the problems involving the complete determination of the extra dimension structure through flight time deviation measurements. This estimation on the size of the extra dimension is much lower than the sensitivity range of tabletop experiments and of those conducted in the LHC. 

Furthermore, it is generally the case that when a higher number of extra dimensions is taken into account the chances of making the phenomenon considered be detectable increases (see Ref. \cite{zwiebach_2004}). The calculations of lightcone fluctuation effects for higher than five dimensional quasiperiodically compactified Kaluza-Klein models are no easy task. However, for each additional extra dimension considered, $\Delta t$ can be expected to be proportional to increasingly negative powers of $\ell_c$, which can result in deviations on the flight time of photons which are more likely to be detected even if the length of the extra dimensions are greater than our current estimative in $(4+1)$.

Our results in a flat background also paves the way for the possibility of a similar analysis for the structure of the extra dimension to be reproduced in more realistic scenarios. In particular, in a cosmological background with a spatially flat $(3+1)$ Friedmann-Lema\^itre-Robertson-Walker metric lightcone fluctuation effects have been analyzed in Ref. \cite{ford1995}. As we deal with cosmological distances and relic gravitons produced in the early Universe are expected to be a potential source of metric fluctuations, we could wonder whether it is possible to generalize the results of Ref. \cite{ford1995} to include a compactified extra dimension, as well as a cosmological constant, and investigate the role played by the scale factor $a(t)$ in determining the structure of the extra dimension by also using the sensitivity of measure devices such as the NIRSpec. For future work we are also interested in searching for observable scenarios on lightcone fluctuations effects by including temperature corrections in the calculations, considering additional extra dimensions or working with curved backgrounds.

\acknowledgments
The author G.A. is financed by the Brazilian agency Coordination for the Improvement of Higher Education Personnel (CAPES) - Finance Code 001. The author H.F.S.M. is partially supported by the Brazilian agency National Council for Scientific and Technological Development (CNPq) under grant No 311031/2020-0.

\appendix 

\section{The Hadamard Function for a scalar field in a flat 5D spacetime with quasiperiodically compactified extra dimension}
\label{apx:D}
Eq. \eqref{fk_5D_genSol} allows us to write the first term on the r.h.s. of Eq.  \eqref{Gzzzz_planeWaveExp} as
\begin{equation}
    D(x,x') = \text{Re}\: \sum_\textbf{k} \frac{e^{-i\omega \Delta t}}{2\omega} \f_\textbf{k}(\textbf{x})\f^{*}_\textbf{k}(\textbf{x}'),
\end{equation}
where $\omega = |\textbf{k}|$, $\textbf{k}= (\textbf{k}_T, n)$, being $\textbf{k}_T = (k_x,k_y,k_z)$, $n$ the quantum number associated with the discretized momentum in the direction of compactification, and $\phi_\textbf{k}(\textbf{x})$ and its Hermitian conjugate are the spatial part of the scalar fields associated with the five-dimensional quasiperiodically compactified spacetime. From Eqs. \eqref{phi_5D}, \eqref{eigValueEq_5D}, and \eqref{momentaSum_5D} we can write
\begin{equation}
    D(x,x') = \text{Re}\: \frac{1}{(2\pi)^3 \ell_c} \sum_{n=-\infty}^\infty \int d^3 \textbf{k}_T \: \frac{e^{-i\omega\Delta t}}{2\omega}e^{i\textbf{k}_T\cdot\Delta \textbf{x}_T + i\frac{2\pi}{\ell_c}(n+\alpha)\Delta w},
\end{equation}
where $\Delta x_T^i = x^i-x'^i$ for $i=1,2,3$, and $\Delta w =  w-w'$. It will be useful to write terms of $\omega$ in the r.h.s. of the integral above as \cite{lightconeCorda}
\begin{equation}
    \frac{e^{-i\omega\Delta t}}{2\omega} = \frac{1}{\sqrt{\pi}}\int_0^\infty ds\: e^{-\omega^2 s^2+\frac{\Delta t^2}{4s^2}}.
\end{equation}
By doing this and performing the integrals over $k_x$, $k_y$ and $k_z$ we obtain
\begin{equation}\label{D_inicial_quasiP}
    D(x,x') = \: \text{Re}\: \frac{1}{8\pi^2 \ell_c} \sum_{n=-\infty}^\infty \int_{0}^{\infty} \frac{ds}{s^3} e^{-\frac{\Delta \textbf{x}_T^2-\Delta t^2}{4s^2} -\left(\frac{2\pi s}{\ell_c}\right)^2(n+\alpha)^2 + \frac{i2\pi(n+\alpha)}{\ell_c}\Delta w}.
\end{equation}
One can use the Jacobi Theta function, namely
\begin{equation}
    \vartheta (u,\mu) = \sum_{n=-\infty}^\infty e^{i\pi n^2 \mu + i2\pi n u},
\end{equation}
to rewrite exponential terms of $n$ and $\alpha$ on \eqref{D_inicial_quasiP} as
\begin{equation}\label{partialSumTheta}
    \sum_{n=-\infty}^\infty e^{-\left(\frac{2\pi s}{\ell_c}\right)^2(n+\alpha)^2 + \frac{i2\pi(n+\alpha)}{\ell_c}\Delta w} =  e^{-\left(\frac{2\pi\alpha s}{\ell_c}\right)^2 + \frac{i2\pi \alpha \Delta w}{\ell_c}}\vartheta\left(i\alpha\frac{4\pi s^2}{\ell_c^2}+\frac{\Delta w}{\ell_c}, i\frac{4\pi s^2}{\ell_c^2}\right).
\end{equation}
The Jacobi theta function has the property \cite{FENG_2014}
\begin{equation}
    \vartheta(u,\mu) = \vartheta \left(\frac{u}{\mu},-\frac{1}{\mu}\right) \frac{e^{-i\pi u^2/\mu}}{(-i\mu)^\frac{1}{2}},
\end{equation}
which we can use to write the sum in \eqref{partialSumTheta} as
\begin{equation}\label{ellipticTheta_transform}
     \sum_{n=-\infty}^\infty e^{-\left(\frac{2\pi s}{\ell_c}\right)^2(n+\alpha)^2 + \frac{i2\pi(n+\alpha)}{\ell_c}\Delta w} = \sum_{n=-\infty}^\infty \frac{\ell_c}{2\pi^\frac{1}{2} s} \: e^{-\frac{(\Delta w + n\ell_c)^2}{4s^2}+i2\pi n\alpha}.
\end{equation}
Substituting Eq. \eqref{ellipticTheta_transform} back in Eq. \eqref{D_inicial_quasiP} and taking the real part one obtains
\begin{equation} \label{D_5D_QP}
    D(x,x') = \frac{1}{8\pi^2} \sum_{n=-\infty}^\infty \frac{\cos(2\pi\alpha n)}{\left[\Delta \textbf{x}_T^2 + (\Delta w + n\ell_c)^2 - \Delta t^2\right]^\frac{3}{2}}.
\end{equation}
Performing these same steps for a free five-dimensional spacetime one finds that the $n=0$ term of Eq. \eqref{D_5D_QP} is exactly the Minkowski contribution. Subtracting such quantity and assuming the propagation limited to the $z$-direction yields the renormalized two-point function
\begin{equation} \label{DR_5D_photonInZ}
    D^R(t,z,t',z') = \frac{1}{8\pi^2} \sideset{}{'}\sum_{n=-\infty}^\infty \frac{\cos(2\pi\alpha n)}{\left[\Delta z^2 + (n\ell_c)^2 - \Delta t^2\right]^\frac{3}{2}},
\end{equation}
where the primed summation indicates that the $n=0$ term has been subtracted.
%
\section{Calculation of $F_{zz}(x,x')$ and $H_{zzzz}(x,x')$}
\label{apx:HeF}
%
Now we turn our attention to finding closed expressions for $F_{zz}(x,x')$ and $H_{zzzz}(x,x')$. From Eqs. \eqref{Fzz_planeWaveExp}, \eqref{kw_5D}, \eqref{phi_5D} and \eqref{eigValueEq_5D} we have
\begin{equation}\label{Fzz_inicialAp}
    F_{zz}(x,x') = - \text{Re}\: \frac{\partial_{\Delta z}^2}{(2\pi)^3 \ell_c} \sum_{n=-\infty}^\infty \int d^3 \textbf{k}_T \: \frac{e^{-i\omega \Delta t}}{2\omega^3} e^{i\textbf{k}_T\cdot\Delta \textbf{x}_T + i\frac{2\pi}{\ell_c}(n+\alpha)\Delta w},
\end{equation}
where $\textbf{k}_T$, $\Delta\textbf{x}_T$, $\Delta w$ and $\omega$ are the same as those mentioned in the previous appendix. It is simple to verify that the fraction involving terms of $\omega$ in the integrand above can be written in terms of the consecutive integrals
\begin{equation}\label{consecInteg_F}
    \frac{e^{-i\omega\Delta t}}{\omega^3} = -\int_0^{\Delta t}dt_2 \int_0^{t_2}dt_1 \frac{e^{-i\omega t_1}}{\omega} - \frac{i\Delta t}{\omega^2} + \frac{1}{\omega^3}.
\end{equation}
Substituting Eq. \eqref{consecInteg_F} back in Eq. \eqref{Fzz_inicialAp} provides
\begin{equation}\label{Fzz_intermedStep}
    F_{zz}(x,x')  = \partial^2_{\Delta z} \int_0^{ \Delta t}dt_2 \int_0^{t_2}dt_1 D(0,z,t_1,z') - \text{Re}\: \frac{\partial_{\Delta z}^2}{(2\pi)^3\ell_c} \sum_{n=-\infty}^\infty \int \frac{d^3 \textbf{k}_T}{2\omega^3} \: e^{i\textbf{k}_T\cdot\Delta \textbf{x}_T + i\frac{2\pi}{\ell_c}(n+\alpha)\Delta w},
\end{equation}
where $D(0,z,t_1,z')$ is given by \eqref{D_5D_QP}. Performing the integrals in the first term in the r.h.s. of Eq. \eqref{Fzz_intermedStep} and using the identity \cite{lightconeCorda}
\begin{equation}\label{integralOmegaFrac}
    \frac{1}{2\omega^{2s}} = \frac{1}{\Gamma(s)}\int_0^\infty d\tau \tau^{2s-1}e^{-\omega^2 \tau^2},
\end{equation}
combined with Eq. \eqref{ellipticTheta_transform} in the second term in the r.h.s. gives us 
\begin{equation} \label{Fzz_5D_QP}
    F_{zz}(t,z,t',z') = \frac{\partial_{\Delta z}^2}{8\pi^2} \sum_{n=-\infty}^\infty \cos(2\pi n\alpha) \frac{\sqrt{\Delta z^2 + (n\ell_c)^2-\Delta t^2}}{\Delta z^2 + (n\ell_c)^2}.
\end{equation}
Once again, the same steps apply to compute the Minkowski contribution $F^{(0)}(t,z,t',z')$, which will be found to coincide exactly with the $n=0$ term of \eqref{Fzz_5D_QP}. So the renormalized $F_{zz}^R$ for a propagation along the $z$-direction is given by
\begin{equation}\label{Fzz_5D_QP_R}
    F^R_{zz}(t,z,t',z') = \frac{\partial_{\Delta z}^2}{8\pi^2} \sideset{}{'}\sum_{n=-\infty}^\infty \cos(2\pi n\alpha) \frac{\sqrt{\Delta z^2 + (n\ell_c)^2-\Delta t^2}}{\Delta z^2 + (n\ell_c)^2}.
\end{equation}

The procedure to obtain $H_{zzzz}(x,x')$ is quite similar. From Eqs. \eqref{Hzzzz_planeWaveExp}, \eqref{kw_5D}, \eqref{phi_5D} and \eqref{eigValueEq_5D} one can write
\begin{equation}\label{Hzzzz_inicialAp}
    H_{zzzz}(x,x') = \text{Re}\: \frac{\partial_{\Delta z}^4}{(2\pi)^3 \ell_c} \sum_{n=-\infty}^\infty \int d^3 \textbf{k}_T \: \frac{e^{-i\omega \Delta t}}{2\omega^5} e^{i\textbf{k}_T\cdot\Delta \textbf{x}_T + i\frac{2\pi}{\ell_c}(n+\alpha)\Delta w}.
\end{equation}
Just as we have done for $F_{zz}(x,x')$, we can write the terms involving $\omega$ in the integrand above as 
\begin{equation}\label{consecInteg_H}
    \frac{e^{-i\omega\Delta t}}{\omega^5} = - \int_0^{\Delta t} dt_2 \int_0^{t_2} dt_1 \frac{e^{-i\omega t_1}}{\omega^3} - \frac{i \Delta t}{\omega^4} + \frac{1}{\omega^5},
\end{equation}
which can be easily verified by performing the integrals in the r.h.s. Substitution of Eq. \eqref{consecInteg_H} back in Eq. \eqref{Hzzzz_inicialAp} yields
\begin{equation}\begin{split}\label{Hzzzz_midAP}
    H_{zzzz}(x,x') & = - \partial_{\Delta z}^2 \int_0^{\Delta t} dt_2 \int_0^{t_2} dt_1 F_{zz}(0,z,t_1,z') + \text{Re}\:\frac{\partial^4_{\Delta z}}{(3\pi)^2 \ell_c} \sum_{n=-\infty}^\infty \int \frac{d^3 \textbf{k}_T}{2\omega^5}\: e^{i\textbf{k}_T\cdot\Delta \textbf{x}_T + i\frac{2\pi}{\ell_c}(n+\alpha)\Delta w} \\
    & = I^{H1}_{zzzz} + I^{H2}_{zzzz},
\end{split}\end{equation}
where $F_{zz}(0,z,t_1,z')$ is given by \eqref{Fzz_5D_QP}. Performing the integrals in the first term in the r.h.s. gives us
\begin{equation}\begin{split}\label{IH1}
    I^{H1}_{zzzz} & = - \frac{\partial^4_{\Delta z}}{8\pi^2} \sum_{n=-\infty}^\infty \cos(2\pi n\alpha) \Biggl\{-\frac{\sqrt{\Delta z^2+(n\ell_c)^2}}{3} + \frac{[2\Delta z^2 + 2(n\ell_c)^2+\Delta t^2]\sqrt{\Delta z^2 + (n\ell_c)^2-\Delta t^2}}{6[\Delta z^2+(n\ell_c)^2]} \\
    & \quad + \frac{\Delta t}{2}\text{arctan}\left[\frac{\Delta t}{\sqrt{\Delta z^2 + (n\ell_c)^2 - \Delta t^2}}\right]\Biggl\}.
\end{split}\end{equation}
On the other hand, using Eqs. \eqref{integralOmegaFrac} and \eqref{ellipticTheta_transform} in the second term in the r.h.s. of \eqref{Hzzzz_midAP} provides
\begin{equation}\label{IH2_div}
    I^{H2}_{zzzz} = \frac{\partial^4_{\Delta z}}{8\pi^2}\frac{2\pi^\frac{1}{2}}{3} \sum_{n=-\infty}^\infty \cos(2\pi n\alpha) \int_0^\infty d\tau e^{-\frac{\Delta z^2+(n\ell_c)^2}{4\tau^2}}.
\end{equation}
Notice that the integral over $\tau$ in Eq. \eqref{IH2_div} does not converge. An easy way out is to simply permute and execute a single derivative in $\Delta z$ to explicitly obtain a term of $\tau^{-2}$, and then proceed to perform the integral. By doing this one obtains
\begin{equation}
    I^{H2}_{zzzz} = \frac{\partial^3_{\Delta z}}{24\pi^2} \sum_{n=-\infty}^\infty \frac{\Delta z}{\sqrt{\Delta z^2 + (n\ell_c)^2}},
\end{equation}
which exactly cancels the first term in the r.h.s. of Eq. \eqref{IH1} upon substitution back in \eqref{Hzzzz_midAP}. Finally, we obtain the $H_{zzzz}(t,z,t',z')$ for a photon propagating in the $z$-direction as
\begin{equation}\begin{split}\label{Hzzzz_5D_QP_R}
    H^R_{zzzz}(t,z,t',z') & = - \frac{\partial^4_{\Delta z}}{8\pi^2} \sideset{}{'}\sum_{n=-\infty}^\infty\cos(2\pi n\alpha) \Biggl\{\frac{[2\Delta z^2 + 2(n\ell_c)^2+\Delta t^2]\sqrt{\Delta z^2 + (n\ell_c)^2-\Delta t^2}}{6[\Delta z^2+(n\ell_c)^2]} \\
    & \quad + \frac{\Delta t}{2}\text{arctan}\left[\frac{\Delta t}{\sqrt{\Delta z^2 + (n\ell_c)^2 - \Delta t^2}}\right]\Biggl\},
\end{split}\end{equation}
where, once again, we replaced the sum $\sum \rightarrow \sum'$ since the term $n=0$ corresponds to the Minkowski contribution, as it can be easily verified by retaking the same steps for the free scalar field in a five-dimensional Minkowski spacetime. 

\bibliographystyle{JHEP-custom}
\end{document}